\documentclass[twocolumn,aps,prb,superscriptaddress]{revtex4-2}
	\usepackage{chngcntr}
	\usepackage{graphicx}
	\usepackage{color}
	\usepackage{amsmath}
	\usepackage{enumitem}
	\usepackage{amssymb}
	\usepackage{cancel}
	\usepackage{ulem}
	\usepackage{multirow}
    \usepackage[colorlinks=true,linkcolor=blue,citecolor=blue]{hyperref}
    
	\newcommand{\be}{\begin{equation}}
		\newcommand{\ee}{\end{equation}}
	
	\newcommand{\bea}{\begin{eqnarray}}
		\newcommand{\eea}{\end{eqnarray}}

	\newcommand{\lb}{\left[}
	\newcommand{\rb}{\right]}
	\newcommand{\lp}{\left(}
	\newcommand{\rp}{\right)}

	\renewcommand{\vec}[1]{{\boldsymbol #1}}

\usepackage[dvipsnames]{xcolor}

	\usepackage{bbold}

\begin{document}

\title{A path to superconductivity via strong short-range repulsion in a spin-polarized band
}
\author{Zhiyu Dong}
\affiliation{Department of Physics and Institute for Quantum Information and Matter, California Institute of Technology, Pasadena, California 91125}
\author{Patrick A. Lee}
\affiliation{Department of Physics, Massachusetts Institute of Technology, Cambridge, MA 02139}

\begin{abstract}
We predict that the spin-polarized electrons in a two-dimensional triangular lattice with strong electron-electron repulsion gives rise to f-wave pairing. The key point is that the first-order interaction, which is usually pair-breaking, vanishes or nearly vanishes in certain f-wave channels due to symmetry constraints. As a result, these f-wave pairing channels are governed by the subleading-order processes which enable pairing when the perturbation theory is controlled. We illustrate this using the Hubbard model on the triangular lattice with on-site and nearest-neighbor repulsion, where we find a $T_c\sim 1\% $ of electron's bandwidth. For a general screened interaction, the same idea works asymptotically, but a third-order calculation is needed to fully determine the strength of f-wave pairing.
\end{abstract}
\maketitle

Electrons strongly repel each other via the Coulomb interaction. In two-dimensional (2D) layered materials, it is possible to screen the Coulomb interaction by adding a metallic plane at distance $d$ away, but the resulting short-range repulsion is still strong. The question is whether a purely repulsive interaction can be shown to give rise to pairing in a theory with an expansion parameter so that the theory is under control. 
 If the answer is positive, a natural follow-up question is whether high transition temperature ($T_c$) can be achieved, because in this case the energy scale is electronic which can be much higher than the Debye scale coming from electron-phonon coupling in conventional BCS theory. Indeed, many believe that the cuprate high-$T_c$ family is driven by electron repulsion via the Mott transition, but a controlled theory is lacking. The goal of this work is to show that a path to superconductivity (SC) using a controlled expansion is possible for the simpler case of a spin-polarized band.

Sixty years ago Kohn and Luttinger (KL) \cite{Kohn1965} provided an answer to the first question. They expanded the BCS kernel up to second order in the coupling constant  and showed  that due to the $2k_F$ singularity, in 3D the second order term gives rise to an attractive channel which overwhelms the first-order repulsion for large enough angular momentum $l$. Unfortunately, $T_c$ is exponentially small in $l^4$  and subsequent extensions \cite{Fay1968, kagan1988possibility,baranov1993enhancement,kagan1991strong,Maiti2013,Kagan2016,Kagan2014,Zanchi1996,Zanchi2000,Raghu2010,fujimoto2025renormalization}, while improving the situation, have yet to demonstrate that the KL mechanism is a promising path to significant $T_c$.

Inspired by discoveries of SC in polarized bands in moire materials\cite{han2025signatureschiralsuperconductivityrhombohedral}, we focus on the question of pairing in a fully polarized band. There have been a number of recent papers on this topic. The analytic work has mostly employed the random phase approximation\cite{chou2024intravalleyspinpolarizedsuperconductivityrhombohedral,geier2024chiraltopologicalsuperconductivityisospin,yang2024topologicalincommensuratefuldeferrelllarkinovchinnikovsuperconductor,qin2024chiralfinitemomentumsuperconductivitytetralayer,jahin2024enhancedkohnluttingertopologicalsuperconductivity,Parra-Martinez2025}, an approach that we have criticized. \cite{dong2026controlled} On the other hand, by exact diagonalization, $f$-wave pairing has been found in a model of electrons with repulsive interactions subject to periodic magnetic flux. \cite{Guerci2025} We would like to understand the generality of this result and whether a controlled path can lead us to find SC pairing out of repulsion. We also mention a recent work on polarized atomic Fermi gas with repulsive ``soft core" repulsion which employs functional renormalization group (FRG) to find $f$-wave pairing.\cite{keles2024pairing} While FRG is not fully controlled, and umklapp is absent in free space, it is encouraging that this work also points to strong  $f$-wave pairing.

The fully spin-polarized problem is simpler because exchange interaction keeps the electrons apart, so that a delta function  interaction does nothing. There is then hope to carry out a perturbation  expansion in the deviation from this trivial limit. The strategy is to find an additional small parameter $\eta$ so that  an effective coupling $g_{\rm eff}=g\eta$ emerges and a controlled expansion is possible, even if the bare coupling $g$ is large. However, this strategy faces two challenges. (1) In a solid, momentum is conserved up to reciprocal vectors $\mathbf{G}$ and   umklapp scatterings involving large momentum transfer in general need to be kept which may not be reduced by $\eta$. (2)
When perturbative expansion is controlled, the first-order contribution dominates. However, the first-order process often has a repulsive sign in all odd-parity pairing channels. This can be shown analytically in a parabolic band with Coulomb interaction \cite{Maiti2013}.
Consequently, in a controlled expansion theory it is usually difficult to get pairing.

In our earlier paper~\cite{dong2026controlled} we indeed found a small  parameter by taking into account the quantum metric effect in the overlap of Bloch functions in the interaction Hamiltonian. 
The current paper extends and improves upon the earlier work 
which assumed a specially tailored interaction [constant for $q<1/d$ and zero otherwise] and an inversion-symmetric Bloch Hamiltonian $H(\mathbf{k}) = H(-\mathbf{k})$. 
We find that these two assumptions 
are unnecessarily restrictive: they remove the entire first-order contribution to the pairing kernel, whereas SC  only requires one pairing channel to evade the first-order repulsion.  

As illustrated in Fig.~1, in this paper we consider an interaction with a smooth momentum dependence. It behaves as $\sim q^2$ at small $q$ up to a momentum scale of $1/d$. 
 \begin{figure}
     \centering
     \includegraphics[width=0.98\linewidth]{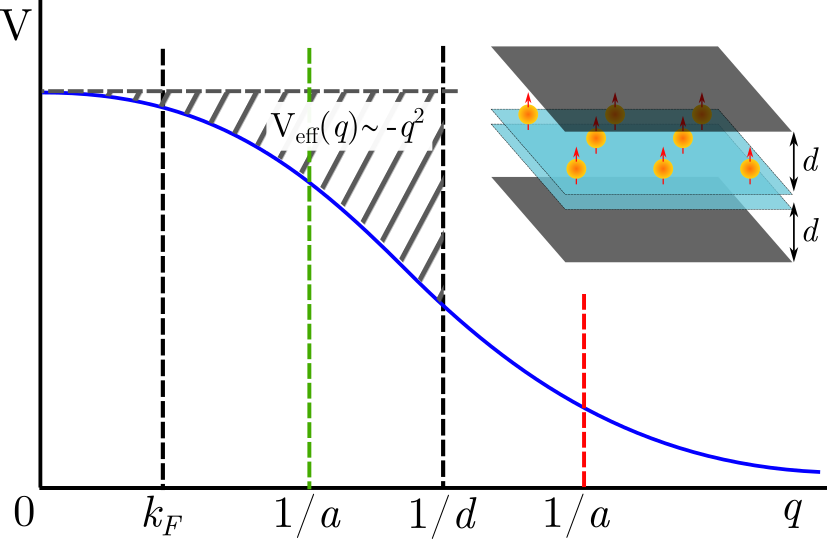}
     \caption{ The inset shows the geometry we propose: a screening plane is placed at a distance $d$ on each side of a spin-polarized 2D sample, which can be either  a triangular crystal or a moire superlattice formed by two layers with  in-plane unit cell size $a$.  The main figure illustrates  the momentum dependence of interaction $V(q)$. The idea is that for small $k_F$ the shaded part of the interaction is small and we can identify a small expansion parameter even if the bare interaction $V(0)$ is large. There are two regimes to consider: the weak-screening regime with $d\gg a$ (red dashed line) and the strong-screening regime with $d\ll a$ (green dashed line). We analyze them separately in Sec.\ref{sec:general analysis}.}
     \label{fig:Veff}
 \end{figure}
The interaction Hamiltonian generally takes the following form 
\be\label{eq:Hint}
H_{\rm int} = \frac{1}{2V}\sum_{\vec q \in BZ}\sum_{\vec G} V_{\vec q+\vec G} \rho_{-\vec q-\vec G} \rho_{\vec q+\vec G}, 
\ee
where $\sum'$ denotes summation within BZ, the density operator $\rho_{\vec q+
\vec G} = \sum'_{\vec k} \Lambda_{\vec k, \vec k+\vec q}^{\vec G}c^{\dagger}_{\vec k+\vec q}c_{\vec k}$, the form factor $\Lambda_{\vec k'\vec k}^{\vec G} = \langle \Psi_{\vec k}| e^{i(\vec k-\vec k'-\vec G)\cdot \vec r}|\Psi_{\vec k'}\rangle = \langle u_{\vec k}| e^{-i\vec G \cdot \vec r}|u_{\vec k'}\rangle $. Here we have used the convention $|u_{\vec k+\vec G}\rangle  = |u_{\vec k}\rangle e^{-i\vec G \cdot \vec r}$ whereas the Bloch wavefunction satisfies $|\Psi_{\vec k+\vec G}\rangle = |\Psi_{\vec k}\rangle$.
We start from the gap equation. Since the first-order pairing interaction $\Gamma^{(1)}$ is frequency independent we can 
carry out summation over Matsubara frequency in the BCS pairing kernel $\sum_m (\omega_m^2+\epsilon_{k'}^2)^{-1} =\tanh (\frac{\epsilon_{\vec k'}}{2T})/\lp 2\epsilon_{\vec k'} \rp  $, and get to the following form of the linearized gap equation with only momentum dependence:
\begin{align}\label{eq:pairing kernel1}
\Delta(\vec k) = \frac{1}{V} \sum_{\vec k'} \frac{\tanh\frac{\epsilon_{\vec k'}}{2T}}{2\epsilon_{\vec k'}} \Gamma^{(1)}(\vec k;\vec k')\Delta(\vec k'), 
\end{align}
where the first-order interaction is given by 
\be\label{eq:Gamma1_VLambdaLambda}
\Gamma^{(1)}(\vec k;\vec k') = -
P_{-}
\sum_{\vec G} V_{\vec k-\vec k'-\vec G} \Lambda_{\vec k'\vec k}^{\vec G}\Lambda_{-\vec k',-\vec k}^{-\vec G}
\ee
where 
$P_{-}$ denotes the odd-odd parity projection operator which is defined as $P_{-} f(k,k') = \frac{1}{4} \lb f(\vec k,\vec k')-f(\vec k,-\vec k')-f(-\vec k,\vec k')+f(-\vec k,-\vec k')\rb $.
The eigenmode $\Delta(\vec k)$ can be classified by irreps of point group $G$, so we label them using the notation $\Delta_{\mathbf{\Gamma}_i}(\vec k)$ where $\mathbf{\Gamma}_i$ is an irrep. As we said, the first-order interaction usually has a repulsive sign (pair-breaking). 
Our goal is to find a channel $\Delta_{\mathbf{\Gamma}_i}$ with a vanishing first-order interaction.

We illustrate this idea with 
an extended Hubbard model on a triangular lattice with nearest-neighbor (NN) repulsion $U_1$. The large onsite repulsion plays no role because of Pauli exclusion. For short-range repulsion, $U_1$ can be small enough that it can be used as an expansion parameter and the umklapp terms mentioned in problem (1) can be included. A controlled expansion indeed yields $f$-wave pairing. The coupling strength can reach $\sim 0.2$ which is strong enough to give a reasonably high $T_c$. In a generic band with a screened Coulomb interaction, we find that for both weak and strong screening regimes, there are definitely f-wave pairing channels that evade first-order repulsion. However, to determine  whether there is an instability towards the f-wave pairing requires a third-order calculation. In comparison, the nearest-neighbor Hubbard model represents a special limit inside the strong-screening regime, where our prediction of f-wave pairing through second-order perturbation theory is reliable.

\subsection{Nearest neighbor Hubbard model}\label{sec:hubbard model}

In this section we study the extended Hubbard model on the triangular lattice where spin-polarized (or spinless) fermions have NN hopping $-t$ ($t>0$), an onsite repulsion $U_0$ and a NN repulsion $U_1$. The on-site repulsion has no effect due to Pauli exclusion.

In momentum space:
\begin{align}
H_{U} &= \sum_{\vec G} \frac{1}{V^4}
\sum_{\vec k \vec k' \vec p \vec p'}'
\frac{V(\vec k'-\vec k 
)}{2}c^\dagger_{\vec p'} c^\dagger_{\vec k'} c_{\vec k} c_{\vec p}  \delta_{\vec p'+\vec k'-\vec k-\vec p,\vec G} ,
\nonumber \\
V(\vec q) &= U_0 + 2 U_1 \sum_j \cos(\vec q \cdot \vec a_j), 
\end{align}
where $\sum'$ represents summation over the Brillouin zone. The argument of $V(\vec k'-\vec k 
)$ can be outside the first BZ, but $V(\vec q)$ is periodic in the extended BZ and is bounded. Note that while the on-site repulsion $U_0$ term can give rise to a spontaneous spin polarization, it has vanishing effects in the spin-polarized phase due to Pauli exclusion. The pairing interaction in the spin-polarized band solely arises from $U_1$. In the expression of $V(q)$, the three NN bond vectors that are summed over are defined as $\vec a_1 = [1,0], \vec a_2 = [-\frac{1}{2}$, $\frac{\sqrt{3}}{2}]$ and $\vec a_3 = [-\frac{1}{2},-\frac{\sqrt{3}}{2}]$. 
At the first order of interaction, the antisymmetrized pairing interaction 
is
\begin{align}
&\Gamma^{(1)}(\vec k; \vec k') = \frac{1}{2} \lp V(\vec k-\vec k') -V(\vec k+\vec k')\rp\\
&= U_1 \sum_{j=1}^{3} \sin(\vec k' \cdot \vec a_j ) \sin(\vec k \cdot \vec a_j )  \label{eq:Gamma1_Hubbard}
\end{align}
The elements of point group $C_{6v}$ permutes the three $k'$-dependent factors $\lbrace \sin(\vec k' \cdot \vec a_j) \rbrace$ ($j=1,2,3$). Therefore, the three factors $\lbrace \sin(\vec k' \cdot \vec a_j) \rbrace$ induces a three-dimensional representation of $C_{6v}$, denoted as $\vec 3$. Using the character table of $C_{6v}$, we find
this 3D representation can be reduced to two irreps: $\vec 3 \cong B_1+E_1$. Here, $E_1$ is the two-dimensional irrep that describes the transformation of an in-plane polar vector under the point group. 
However, the point group $C_{6v}$ has three types of odd-parity pairing channels in total, labeled by three odd-parity irreps: $E_1$ (two degenerate $p$ waves), $B_1$ ( $f$ wave with lobes along the NN bonds) and $B_2$ ( $f$ wave with nodes along the NN bonds).
As a result, the first-order interaction has to vanish in the pairing channel $B_2$ because $\Delta_{B_2}(\vec k')$ and the $\vec k'$-dependence of $\Gamma^{(1)}(\vec k;\vec k')$ transform differently under the point group. In fact, this result can be directly understood by looking at the behavior of gap functions under the three $\sigma_v$ mirror operators which lie along three $120^\circ$ crystal axes $\vec a_1$, $\vec a_2$ and $\vec a_3$: The gap function in $B_2$ channel is odd under all three $\sigma_v$ mirrors whereas $\sin(\vec k' \cdot \vec a_j)$ 
is even under the $\sigma_v$ mirror that lies along $\vec a_j$. Therefore, the overlap $ \int d\vec k' \sin(\vec k' \cdot \vec a_j) \Delta_{B_2}(\vec k')$ vanishes for any $j=1,2,3$, so the first-order pairing interaction indeed vanishes in
the $B_2$ channel. We note parenthetically that, introducing a next-nearest-neighbor (NNN) repulsion would give rise to a finite first-order interaction in $B_2$ channel\footnote{This is because the NNN terms take the form of $\sum_{b_j} \sin(\vec k \cdot \vec b_j)$ where $\vec b_j$ with $j=1,2,3$ are the three $120^\circ$ vectors connecting NNN. The representation induced by $\lbrace \vec b_j\rbrace$'s contains irrep $B_2$. }. Therefore, for our purpose, we indeed need   the NNN repulsion to be small compared with the NN one.
\begin{figure}
    \centering
    \includegraphics[width=0.98\linewidth]{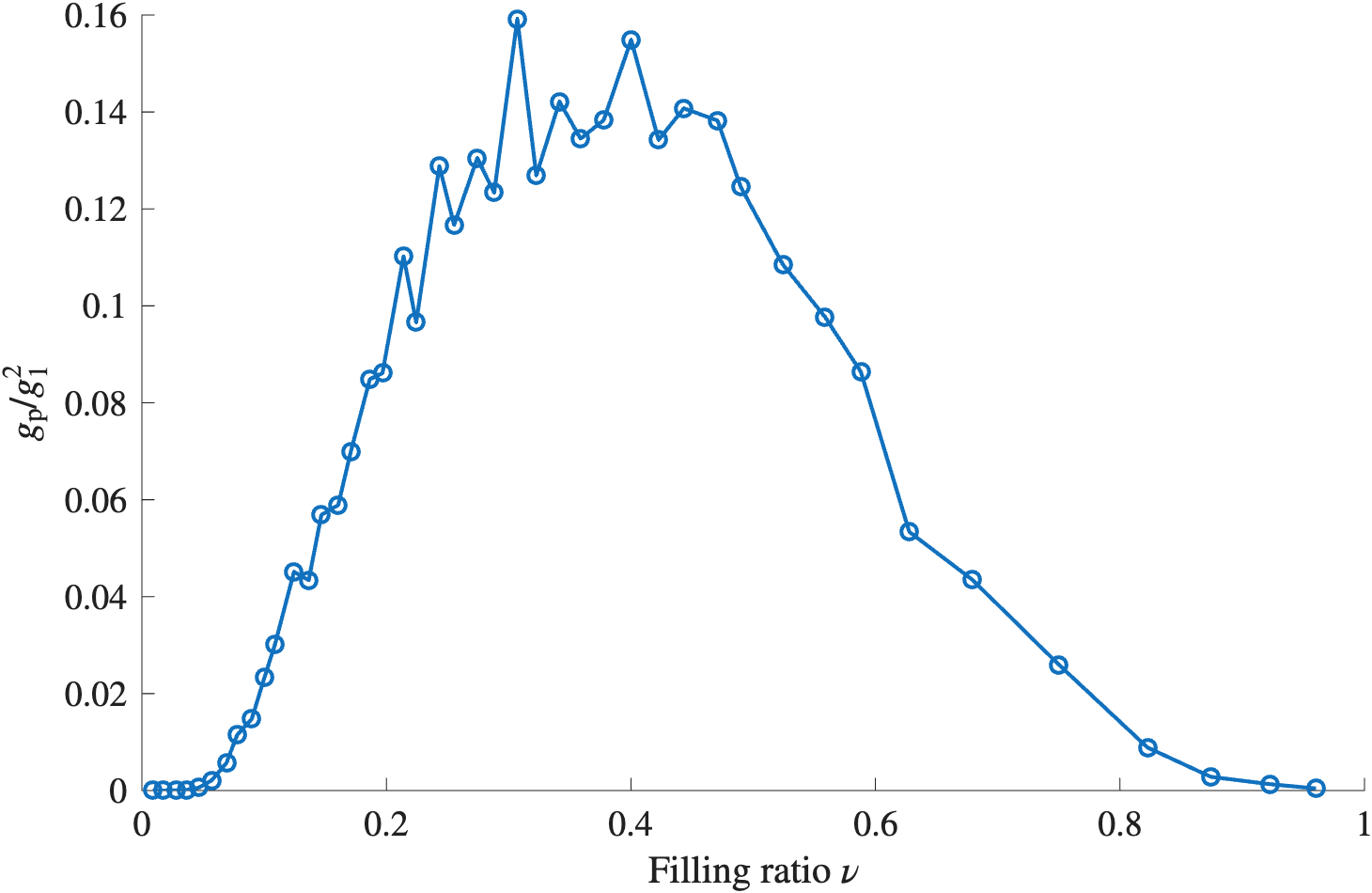}
    \caption{The strength of superconductivity in the nearest-neighbor Hubbard model. 
    We expand in the dimensionless coupling constant $g_1=6U_1A_{\rm uc}\nu_0 $, where $A_{\rm uc}$ is the area of a unit cell, $\nu_0$ is the density of states on the Fermi surface. The first-order contribution vanishes and we compute  the strength of the pairing interaction $g_p$ in the $B_2$ channel ($f$ wave with nodes along the NN bonds)  up to second order in $g_1$. The y-axis shows  $g_p$ divided by $g_{1}^2$.
    \label{fig:Hubbard model}}
\end{figure}

We confirm this picture by  numerically computing the pairing strength to second order in the coupling constant $U_1$. Details are shown in the Appendix. We define the dimensionless coupling $g_1=6U_1A_{\rm uc}\nu_0 $, where $A_{\rm uc}$ is the area of a unit cell, $\nu_0$ is the density of states on the Fermi surface which depends on filling. Unlike the discussion in the introduction, for the Hubbard model the notation $g_{\rm eff}=g \eta$ does not apply because there is no bare coupling and $U_1$ is already the difference compared with a purely onsite repulsion $U_0$. Therefore we are simply expanding in powers of  
$U_1 \propto  g_1$. The strongest pairing channel is  $B_2$ and its strength is shown in Fig.~\ref{fig:Hubbard model} in units of $g_1^2$. Details of how the coupling strength is extracted is  shown in Appendix A.  The pairing strength can reach a value of $\sim 0.2$ at a filling ratio of about 0.4 if we set $g_1=1$. For a system with $\epsilon_F\sim 1\rm{eV}$, this amounts to a $T_c$ of a hundred K. 

The numerics in Fig.\ref{fig:Hubbard model} also show a relative small f-wave pairing interaction at dilute regimes (dilute electrons $\nu\ll 1$ or dilute holes $1-\nu \ll 1$). This is reasonable as the total pairing interaction in the f-wave channel is $\Gamma(\vec k;\vec k')\sim (k_x +ik_y)^3(k'_x +ik'_y)^3 \sim O\lp k_F^6a^6 \rp$ to the leading order of $k_F a$ which is small in dilute regimes. As discussed later, in this regime, the expansion is governed by the smallness of $k_Fa$ and it is necessary to go to third order in perturbation theory to get a reliable result.

Next, we show that the same argument applies to systems with $C_{3v}$ symmetry. As an example, we consider a triangular lattice with different on-site potential on the ABC sublattices whereas the nearest neighbor repulsion is identical on each bond. In this setting, Eq.\eqref{eq:Gamma1_Hubbard} remains valid, only the symmetry analysis below is modified as follows. 
The elements of point group $C_{3v}$ permutes the three $k'$-dependent factors $\lbrace \sin(\vec k' \cdot \vec a_j) \rbrace$ ($j=1,2,3$). Therefore, the three factors $\lbrace \sin(\vec k' \cdot \vec a_j) \rbrace$ induces a three-dimensional representation of $C_{3v}$, denoted as $\vec 3'$. The character of this representation is $\chi_{\vec 3'}: (E, 2C_3,3\sigma_v) = (3,0,1)$ Using the character table of $C_{3v}$, we find $\vec 3' \cong A_2 \oplus E$. The remaining irrep is the trivial irrep $A_1$. For odd-parity pairing, $A_1$ corresponds to an f-wave pairing. As a result, we conclude that, in a $C_{3v}$ system, there still exists an f-wave pairing channel that evades the first-order pairing interaction. On the other hand, these considerations do not work in the square lattice considered in ref.~\cite{dong2026controlled}, because  for $C_4$ symmetry there is only one odd-parity irrep. 

\subsection{General analysis}\label{sec:general analysis}

It is useful to consider two regimes  as shown in Fig.~1 : the weak-screening regime $d\gg a$ and the strong-screening regime $d\ll a$. Below we analyze them separately.

a) Weak screening $d\gg a$:
In this regime, $V_G \ll V_0$, so the umklapp processes can be safely ignored, and there is only $G=0$ term in the Hamiltonian Eq.\eqref{eq:Hint}. In Ref.\cite{Dong2025} we considered the case where $V(q)$ is completely flat up to $q_*=1/d$. In this case, if the electron wavefunction are just plane waves, the direct and exchange interaction exactly cancel. When the wavefunction is not just a plane wave, there will be a nonvanishing effective interaction arising from the quantum geometry which makes the cancellation imperfect. The form of pairing interaction given in Eq.~\eqref{eq:Gamma1_VLambdaLambda} can be seen by first expanding the form factor to leading order of $k$ and $k'$. Due to the constraint of three-fold rotation symmetry, $\Lambda_{\vec k,\vec k'}^0\Lambda_{-\vec k,-\vec k'}^0 = 1 - \xi^2(k^2+ k'^2) - \lambda^2 \vec k \cdot \vec k' $, where $\xi$ and $\lambda$ are characteristic length scales that are obtained by taking gradients of $\Lambda_{kk'}^0$ in momentum space. As a result, the antisymmetrized pairing interaction $\Gamma^{(1)}(\vec k ;\vec k')$ is proportional to $\lambda^2 \vec k \cdot \vec k'$.

For a generic screened interaction, the interaction's q-dependence gives rise to another contribution to Eq.~\eqref{eq:Gamma1_VLambdaLambda}. At leading order of $q$, $V(q) = V_0 - \alpha q^2 d^2 + O(q^4)$, where $\alpha$ is an order-1 number depending on details. As a result, at leading order of $k$ and $k'$, the contribution of $V_q$ to first-order pairing interaction is proportional to $(\vec k-\vec k')^2d^2$, which becomes proportional to $\vec k \cdot \vec k' d^2$ after antisymmetrizing with respect to $\vec k$ and $\vec k'$.

Putting these two contribution together, we find that the total pairing interaction is still proportional to $\vec k\cdot \vec k'$. This is proportional to $k_F^2$ which can serve as a small expansion parameter for low density. Indeed, we identify the effective coupling 
$g_{\rm eff} = g_0 \eta$, where $g_0 = \nu_0 V_0$, and the small parameter $\eta$ is governed by the larger one of two contributions, i.e., $\eta  = (k_F \max(d,\lambda) )^2$.
This $\vec k \cdot \vec k'$ form of $\Gamma^{(1)}$ leads to an ideal situation for f-wave pairing because  $\Gamma^{(1)}$ transform as vectors, thus only has a finite projection in the $p$-wave pairing channels. 
In conclusion, it looks promising to achieve pairing in $f$-wave channels in this regime




However, there is a caveat: as the effective interaction strength is $g_{\rm eff}=g_0 \eta\sim k_F^2$, the second-order diagrams are no longer the leading contribution to f-wave pairing. This is due to an extra constraint on form of f-wave pairing interaction imposed by symmetry. To the leading order of $k_F$, the f-wave gap's leading momentum dependence is $\Delta(k)\sim (k_x \pm ik_y)^3$. Therefore,
the leading k-dependence of the total interaction that gives rise to f-wave pairing behaves as $\Gamma_{f}(\vec k;\vec k')\sim (k_x \pm ik_y)^3(k'_x \mp ik'_y)^3 \sim O\lp k_F^6 \rp$ in order to satisfy the gap equation.
For small $k_F$, the f-wave pairing interaction is $\sim \eta^3$ for any order in $g_0$. In particular, 
the second-order diagrams' contribution is $\sim g_0^2 \eta^3$. This is of order  $g_{\rm eff}^2 \eta  $. On the other hand, the third-order diagrams' contribution is $g_{\rm eff}^3 \sim g_0^3 \eta^3$. As $g_0\gg 1$ when $g_{\rm eff}\sim 1$, we see that the third-order contribution to f-wave channel is larger than the second-order one and is the dominant contribution in this perturbative expansion. In principle, the small expansion parameter we have identified $g_{\rm eff} = g_0 \eta$ is still operational and a controlled expansion is possible, but we need to go to the third order in $g_{\rm eff}$ which is a daunting task.

b) Strong screening $d\ll a$: In this regime, we need to account for umklapps. We know that for a contact interaction $V=V_0$, no matter what wavefunction is, its effect always strictly vanishes in spin-polarized electrons when all umklapps are accounted for. Therefore, the effective part of the interaction is its difference from contact interaction: $V_{\rm eff}(q) = V(q) - V_0 = -\alpha q^2 d^2 + ...$. Projecting the effective interaction onto the band yields
\be
H_{\rm eff} = -\alpha \sum_{\vec G}\frac{1}{V^3}\sum_{\vec k \vec p \vec q} (\vec q +\vec G)^2 d^2  \Lambda_{\vec k+\vec q,\vec k}^{\vec G} \Lambda_{\vec p-\vec q, \vec p}^{\vec G}
c^{\dagger}_{\vec p-\vec q}c^{\dagger}_{\vec k+\vec q}c_{\vec k} c_{\vec p}
\ee
Therefore, the small parameter $\eta$ which, as a reminder, is defined as the ratio between the effective interaction $g_{\rm eff}$ and the bare one $g_0$, contains the contribution from non-umklapp processes (denoted as $\eta_0$) and finite umklapp processes (denoted as $\eta_1$), i.e. $\eta = \eta_0 + \eta_1$. Here $\eta_0$ and $\eta_1$ are given by
\be\label{eq:Heff umklapp}
\eta_0 = k_F^2 d^2  \langle\Lambda^{0}\rangle_{k_F}^2 \sim k_F^2 d^2
,\quad \eta_1 = \sum_{\vec G \neq 0}\vec G^2 d^2  \langle\Lambda^{\vec G}\rangle_{k_F}^2.
\ee
Here $\langle\Lambda^{\vec G}\rangle_{k_F}$ represents average of $\Lambda^{\vec G}_{\vec k,\vec k'}$ on the Fermi surface. Whether $\eta_0$ or $\eta_1$ dominates depends on how fast $\Lambda^{\vec G}$ decays with $|\vec G|$, which is determined by the structure of wavefunction $u_{\vec k}(r)$. 

For concreteness, we illustrate using a Gaussian-orbital model as an example. In this model, atomic orbital is a Gaussian function $\phi_{\vec R}(r) = \sqrt{\frac{2}{\pi r_0^2}}\exp(-(\vec r-\vec R)^2/r_0^2)$. In this model, the cell-periodic part of Bloch wavefunction is given by $|u_{\vec k}\rangle =\frac{1}{\sqrt{N S(\vec k)}}\sum_{\vec R} e^{-i\vec k \cdot (\vec r - \vec R)}\,|\phi_{\vec R}\rangle ,
$ where the normalization factor $S(\vec k) = \sum_{\vec R} e^{i\vec k \cdot \vec R} \langle \phi_0 |\phi_{\vec R}\rangle $.
The periodic function $u_{\vec k}(\vec r)$ can be decomposed in terms of plane-wave harmonics: $u_{\vec k}(\vec r) = \frac{1}{\sqrt{A_{\rm uc}}} \sum_{\vec G} C_{\vec k}(\vec G)e^{i\vec G \cdot \vec r}$. For the Gaussian orbital, $C_{\vec k}(\vec G)$ with $|\vec G|>1/r_0$ are all negligible. Consequently, the form factors $\Lambda^G$'s are small when $|\vec G|>1/r_0$. Based on this, we can roughly estimate through dimension analysis that $\eta_1 \sim d^2/r_0^2$. Therefore, in this example the competition between $\eta_0$ and $\eta_1$ is determined by the competition between $k_F$ and $1/r_0$. 

Based on this example we expect that, in a general setting with more complicated form of the Wannier orbital, $\eta_1$ will be determined by some length scale $\ell$ characterizing the structure of Bloch wavefunction inside each unit cell, similar to the role of $r_0$ in the Gaussian-orbital model. When $\ell\gg a$, $\eta_0 >\eta_1$, the system is dominated by non-umklapp processes. When $\ell \ll a$, $\eta_0 <\eta_1$, the system is dominated by umklapp processes. The behavior of pairing interaction in non-umklapp-dominated regime and umklapp-dominated regime are distinct. Below we analyze the two regimes separately.
 
(1) In the non-umklapp-dominated regime, Eq.\eqref{eq:Heff umklapp} shows that the interaction is quadratic in momentum transfer $q$, which is similar to the situation in our analysis of weak-screening regime. Therefore, the first-order interaction only contribute in p-wave channels and vanishes in f-wave channels. This lifts the main obstruction of achieving f-wave pairing and makes it promising.
However, to fully determine the strength of f-wave pairing, our second order calculation is not enough. The reason is Eq.\eqref{eq:Heff umklapp} shows that, upon neglecting umklapps, the effective interaction is governed by a small parameter $\eta\sim k_F^2 d^2 $. Therefore, we again encounter the same issue as in the regime of $d\gg a$. As a reminder, the issue is that third-order diagrams might have a larger contribution to the f-wave pairing interaction than second-order diagrams. The third order diagrams are computationally expensive and are beyond the scope of this paper.

(2) In the umklapp-dominated regime, the interaction is no longer proportional to $q^2$. As a result, there is no generic reason for first-order pairing interaction to vanish  or approximately vanish in all f-wave channels, thus cannot generally achieve an f-wave SC through a controlled expansion
. However, in this regime, there is one special limit where an f-wave pairing is achieved, which is when the atomic orbital becomes localized (tight-binding). In this limit ($d\ll a$ and tight-binding orbital), the model asymptotically restores the NN Hubbard model we studied in Sec.\ref{sec:hubbard model}. As a reminder, the pairing interaction in one f-wave channel exactly vanishes in this case, without requiring $k_F$ to be small. Therefore in NN Hubbard model, although dilute-carrier regime the small $k_F$ requires $g_{\rm eff}\sim (k_Fd)^2$ and makes third-order diagram non-negligible, our second-order result for $O(1)$-doping regime is still reliable.


The appearance of the low density parameter $k_F d$ has been discussed for unpolarized Fermi gas.\cite{Fay1968, kagan1988possibility}. The physical origin is that in the long-wavelength limit, scattering is isotropic, so that the repulsive term contributes only to the $s$-wave  channel to the first order. \cite{ kagan1988possibility}. In this case the leading pairing channel is $p$-wave and second-order perturbation theory suffices. 
Another difference is that  in a solid the umklapp terms need to be considered.

In summary, our work suggests a search for superconductivity where a layered metal with a relatively isolated spin-polarized band is placed between two screening planes. 
The moire structures produced by twisting or lattice mismatch in van de Waals materials such as TMDs may be a promising place to start\cite{Andrei2021}. Indeed  spin-polarized states have been observed in twisted WSe$_2$ \cite{knuppel2025correlated}, and Wigner crystals have been seen in WSe$_2$/WS$_2$ hetero moire structures.\cite{li2021imaging} 
The screening gates will truncate the long range part of the Coulomb repulsion and destabilize the Wigner crystal which is a competing state. The transition temperature will be low because the energy scale is low in moire systems, but polarization may be driven with an external Zeeman field if it is not opposed by spin-orbit coupling. On  the other hand, our Hubbard model result may encourage searches in other systems with higher energy scales where polarization may be achieved due to coupling to ferromagnetically aligned local moments. We note that the superconductivity found in spin-polarized phase in rhombohedral graphene\cite{han2025signatureschiralsuperconductivityrhombohedral} might also be related to our theory, as it features a single electron flavor, a strong short-range repulsion and a $C_3$ symmetry. However, this system hosts more complicated orders such as Wigner crystal phases, fractional quantum Hall states, which might be related to SC but are not what our model tries to capture. Therefore, we leave it as an open question.



We thank Liang Fu for insightful discussions and Andrey Chubukov,  Kin Fai Mak and Qianhui Shi for helpful comments. Z. D. acknowledges support from the Gordon and Betty Moore Foundation’s EPiQS Initiative, Grant GBMF8682. P.A.L. acknowledges support from DOE (USA) office of Basic Sciences Grant No. DE-FG02-03ER46076.

\bibliography{ref}

\newpage

\appendix

\section{Extracting pairing interaction $g_p$}\label{appendix:A}
\begin{figure}
    \centering
    \includegraphics[width=0.98\linewidth]{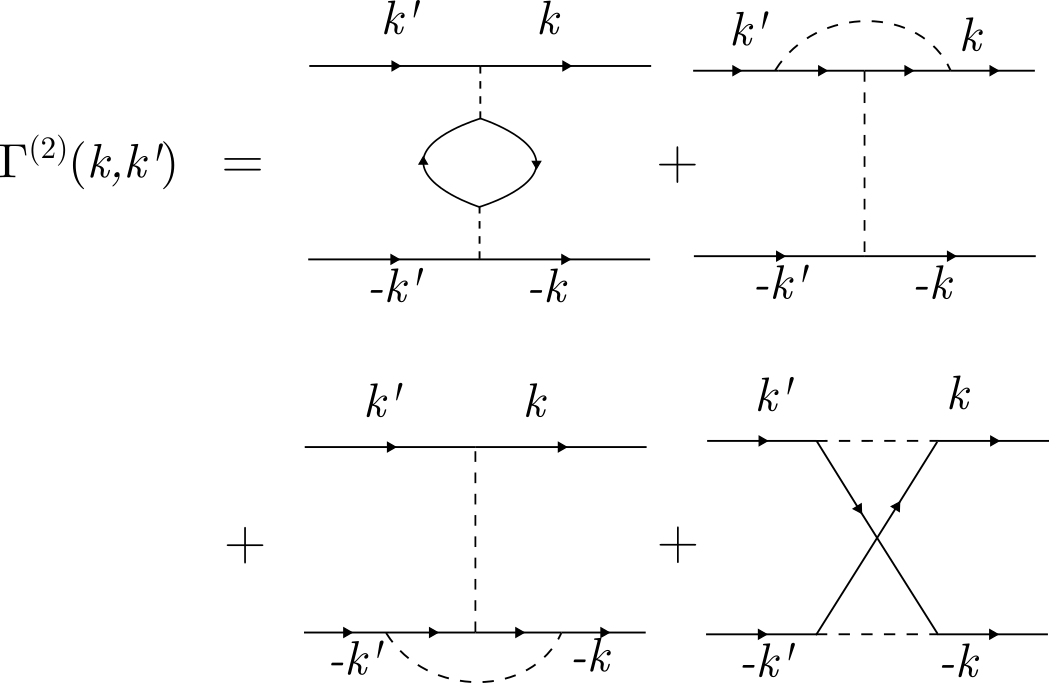}
    \caption{Feynman diagrams for pairing interaction at second order.}
    \label{fig:2nd diagram}
\end{figure}
In this appendix, we explain how we numerically extract the pairing interaction strength $g_p$. We focus on the NN Hubbard model. 
We numerically diagonalize the kernel in BCS gap equation 
\be
\Delta(\vec k,\omega) = \frac{1}{V}\sum_{\vec k' \omega'} \Gamma(\vec k,\omega;\vec k',\omega')G(-\vec k', -\omega') G(\vec k',\omega')\Delta(\vec k',\omega')
\ee
Here we calculate the pairing interaction $\Gamma$ up to second order, which includes the four diagrams shown in Fig.\ref{fig:2nd diagram}. 
To solve the gap equation, we make two approximations: 
\begin{enumerate}
    \item We neglect the self energy in $G(\vec k,\omega)$. We expect the self-energy merely alters the dispersion and will not essentially change our main conclusion. 
    \item  We ignore the frequency dependence in $\Gamma(\vec k,\omega;\vec k',\omega')$, replacing it with the $\omega=\omega'=0$ value of the pairing interaction, $\Gamma(\vec k;\vec k')$. This is exact at first order, but is an approximation for second order. This approximation is acceptable because for weak coupling, pairing is limited to the vicinity of the Fermi surface and  the low frequency pairing interaction correctly describes the leading logarithmic singularity in the pairing channel. The frequency dependence of $\Gamma(\vec k,\omega; \vec k',\omega')$ merely sets the bandwidth of the pairing interaction, which is expected to be of order $\epsilon_F$. Therefore, we expect this approximation will  not affect the coefficient of the BCS logarithm and therefore will not qualitatively change our conclusion.
\end{enumerate}

\begin{figure}
    \centering
\includegraphics[width=0.98\linewidth]{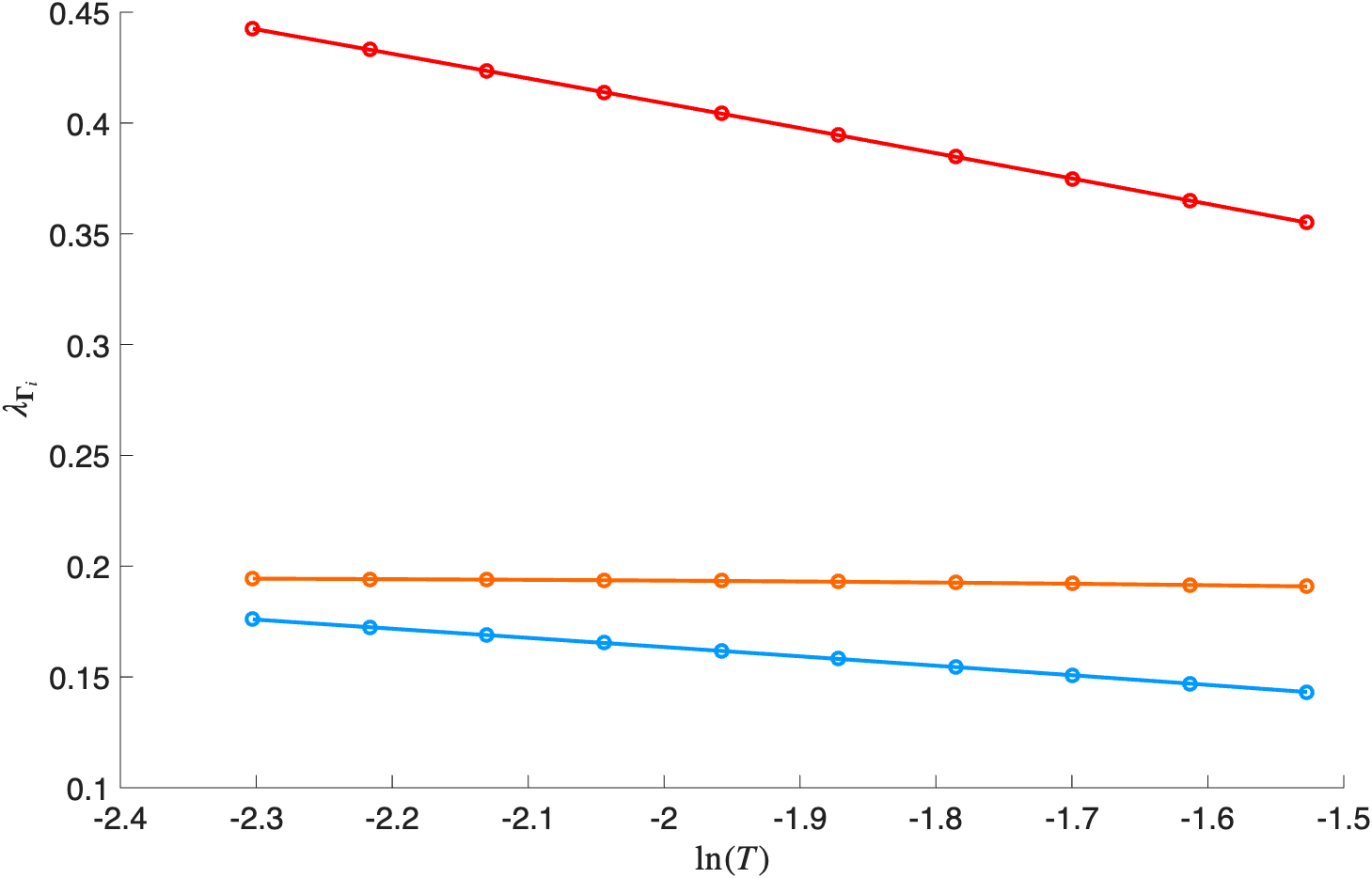}
    \caption{The eigenvalues of pairing kernels in the four leading (attractive) pairing channels as a function of temperature $T$ in the NN Hubbard model, measured at $g_{\rm eff}=1$ and filling $\nu= 0.3$. The red lines represents the $B_2$ (f-wave) channels in which the 1st order repulsion vanishes, whereas the doubly degenerate orange line corresponds to the two degenerate p-wave channels, the cyan line corresponds to the $B_1$ channels (f-wave). At sufficiently low $T$ the leading eigenvalues of two f-wave modes are linear in $\log T$. The eigenvalues in two p-wave channels has a very small slope in $\log T$ because the gap functions in these channels are sign-changing across the Fermi surface (or at least have a strong radial dependence) to evade the non-negligible first-order repulsion. These p-waves are not standard BCS pairing.}
    \label{fig:logT}
\end{figure}

Under these two approximations, the BCS gap equation at finite temperature $T$ becomes
\be
\Delta(\vec k) = \frac{1}{V}\sum_{\vec k'} K(\vec k;\vec k')\Delta(\vec k')
\ee
where $K(\vec k;\vec k') = \frac{\tanh\frac{\epsilon_{\vec k'}}{2T}}{2\epsilon_{\vec k'}} \Gamma(\vec k;\vec k')$ . We solve this equation by numerically diagonalizing the kernal matrix $K(\vec k; \vec k')$ and obtaining the eigenvalues $\lambda_{\mathbf{\Gamma}_i}$'s in all pairing channels $\mathbf{\Gamma}_i$.
We repeat this at each value of $T$ to get $\lambda_{\mathbf{\Gamma}_i}$ as a function of $T$. We verify that  $\lambda_{\mathbf{\Gamma}_i}(T)$ scales linearly with $\log T$ at sufficiently low $T$ (see Fig.\ref{fig:logT}), in agreement with BCS theory.  We extract the strength of the pairing interaction from the slope $g^{\mathbf{\Gamma}_i}_{\rm p} = \partial \lambda_{\mathbf{\Gamma}_i}/ \partial \log T$. Note that from our numerics we often find nearly-non-BCS pairing channels where the gap nearly vanishes at the Fermi surface. This behavior occurs because pairing in $E_1$ and $B_1$ channels sometimes prefer to change sign near Fermi surface to avoid the first-order repulsion. As a result, as seen in Fig.\ref{fig:logT} some of these channels can have much smaller slopes in $\log T$
. In channels with tiny slopes, the contribution to pairing interaction is weak on the Fermi surface, whereas the contribution from states away from Fermi surface can be larger.  Similarly in the case $d > a$, there may be pair excitations that contribute to Eq.~\eqref{eq:pairing kernel1} involving states far from the Fermi energy. This contribution is not $\log T$-divergent and will be ignored.

\end{document}